\input harvmac
\noblackbox
\def\npb#1#2#3{{\it Nucl.\ Phys.} {\bf B#1} (19#2) #3}
\def\pla#1#2#3{{\it Phys.\ Lett.} {\bf A#1} (19#2) #3}
\def\plb#1#2#3{{\it Phys.\ Lett.} {\bf B#1} (19#2) #3}
\def\prl#1#2#3{{\it Phys.\ Rev.\ Lett.} {\bf #1} (19#2) #3}
\def\pra#1#2#3{{\it Phys.\ Rev.} {\bf A#1} (19#2) #3}

\def\prd#1#2#3{{\it Phys.\ Rev.} {\bf D#1} (19#2) #3}

\def\mpla#1#2#3{{\it Mod.\ Phys.\ Lett.} {\bf A#1} (19#2) #3}
\def\mplb#1#2#3{{\it Mod.\ Phys.\ Lett.} {\bf B#1} (19#2) #3}

\def\cmp#1#2#3{{\it Commun.\ Math.\ Phys.} {\bf #1} (19#2) #3}
\def\jgp#1#2#3{{\it J. Geom.\ Phys.} {\bf #1} (19#2) #3}

\def\jpq#1#2#3{{\it J. Physique} {\bf #1} (19#2) #3}
\def\jpql#1#2#3{{\it J. Physique Lett.} {\bf #1} (19#2) L-#3}
\def\jpqII#1#2#3{{\it J. Phys.\ II France} {\bf #1} (19#2) #3}
\def\phsca#1#2#3{{\it Physica} {\bf #1D} (19#2) #3}

\def\endli{\hfill\break}
\def\frac#1#2{{#1 \over #2}}

\def\p{\partial}
\def\sdet{{\rm sdet}\,}
\def\str{{\rm str}\,}
\def\semi{\subset\kern-1em\times\;}
\def\bar#1{\overline{#1}}

                   \def\CL{{\cal L}}
                   \def\CN{{\cal N}}
\def\CO{{\cal O}}

\def\CZ{{\cal Z}}

\def\R{{\bf R}}                     \def\S{{\bf S}}

\def\AdS{{\rm AdS}}
\def\QCD{{\rm QCD}}
\def\OSp{OSp\,}
\Title{\vbox{\baselineskip12pt
\hbox{hep-th/9811028}
\hbox{CALT-68-2199}}}
{\centerline{On QCD String Theory and AdS Dynamics}}
\medskip\bigskip
\centerline{Petr Ho\v rava}
\bigskip
\centerline{\it California Institute of Technology, Pasadena, CA 91125, USA}
\centerline{\tt horava@theory.caltech.edu}
\centerline{\it and}
\centerline{\it Department of Physics and Astronomy, Rutgers University}
\centerline{\it Piscataway, NJ 08855, USA}
\baselineskip18pt
\medskip\bigskip\medskip\bigskip\medskip
\baselineskip16pt
The AdS/CFT correspondence of elementary string theory has been recently 
suggested as a ``microscopic'' approach to QCD string theory in various 
dimensions.  We use the microscopic theory to show that the ultraviolet 
regime on the string world-sheet is mapped to the ultraviolet effects in 
QCD.  In the case of $\QCD_2$, a world-sheet path integral representation of 
$\QCD$ strings is known, in terms of a topological rigid 
string theory whose world-sheet supersymmetry is reminiscent of Parisi-Sourlas 
supersymmetry.  We conjecture that the supersymmetric rigid string theory is 
dual to the elementary Type IIB string theory in the singular AdS background 
that corresponds to the large-$N$ limit of $\QCD_2$.  We also generalize the 
rigid string with world-sheet Parisi-Sourlas supersymmetry to dimensions 
greater than two, and argue that the theory is asymptotically free, a non-zero 
string tension is generated dynamically through dimensional transmutation, and 
the theory is topological only asymptotically in the ultraviolet.  

\Date{October 1998}
\nref\maldacon{J.M. Maldacena, ``The Large N Limit of Superconformal Field 
Theories and Supergravity,'' hep-th/9711200.}
\nref\ewadsholo{E. Witten, ``Anti-de Sitter Space and Holography,'' 
hep-th/9802150.}
\nref\gkp{S.S. Gubser, I.R. Klebanov and A.M. Polyakov, ``Gauge Theory 
Correlators from Non-Critical String Theory,'' \plb{428}{98}{105}, 
hep-th/9802109.}
\nref\ewtherm{E. Witten, ``Anti-de~Sitter Space, Thermal Phase Transition, and 
Confinement in Gauge Theories,'' hep-th/9803131.}
\nref\joereview{J. Polchinski, ``Strings and QCD?,'' hep-th/9211045.}
\nref\ampconf{A.M. Polyakov, ``String Theory and Quark Confinement,'' {\it 
Nucl.\ Phys.\ Proc.\ Suppl.} {\bf 68} (1998) 1, hep-th/9711002; ``Confining 
Strings,'' hep-th/9607049.}
\nref\grosso{D.J. Gross and H. Ooguri, ``Aspects of Large N Gauge Theory 
Dynamics as Seen by String Theory,'' hep-th/9805129.}
\nref\ooguri{C. Cs\'aki, H. Ooguri, Y. Oz and J. Terning, ``Glueball Mass 
Spectrum from Supergravity,'' hep-th/9806021\endli
H. Ooguri, H. Robins and J. Tannenhauser, ``Glueballs and Their Kaluza-Klein 
Cousins,'' hep-th/9806171\endli
R. de Mello Koch, A. Jevicki, M. Mihailescu and J. Nunez, ``Evaluation of 
Glueball Masses from Supergravity,'' hep-th/9806125.}
\nref\miaoli{M. Li, ``'t~Hooft Vortices and Phases of Large N Gauge 
Theory,'' hep-th/9804175; ``Evidence for Large N Phase Transition in 
$N=4$ Super Yang-Mills Theory at Finite Temperature,'' hep-th/9807196.}
\nref\trst{P. Ho\v rava, ``Topological Strings and QCD in Two Dimensions,'' 
in: ``Quantum Field Theory and String Theory,'' Proceedings of the Carg\`ese 
Workshop on ``New Developments in String Theory, Conformal Models, and 
Topological Field Theory,'' May 1993; eds: L. Baulieu et al.\ (Plenum Press, 
1995), hep-th/9311156.} 
\nref\cmr{S. Cordes, G. Moore and S. Ramgoolam, ``Large-$N$ 2D Yang-Mills 
Theory and Topological String Theory,'' \cmp{185}{97}{543}, hep-th/9402107.}
\nref\srst{P. Ho\v rava, ``Topological Rigid String Theory and Two-Dimensional 
QCD,'' \npb{463}{96}{238}, hep-th/9507060.}
\nref\lsew{L. Susskind and E. Witten, ``The Holographic Bound in 
Anti-de~Sitter Space,'' hep-th/9805114.}
\nref\apjp{A.W. Peet and J. Polchinski, ``UV/IR Relations in AdS Dynamics,'' 
hep-th/9809022.}
\nref\grta{D.J. Gross, ``Two Dimensional QCD as a String Theory,'' 
\npb{400}{93}{161}, hep-th/9212149\endli
D.J. Gross and W. Taylor, ``Two Dimensional QCD is a String Theory,'' 
\npb{400}{93}{181}, hep-th/9301068; ``Twists and Wilson Loops in the String 
Theory of Two Dimensional QCD,'' \npb{403}{93}{395}, hep-th/9303046.}
\nref\migrus{A. Migdal, ``Recursion Equations in Gauge Field Theories,'' 
{\it Zh.\ Eksp.\ Theor.\ Fiz.} {\bf 69} (1975) 810\endli
B.Ye.\ Rusakov, ``Loop Averages and Partition Functions in $U(N)$ Gauge 
Theory in Two Dimensional Manifolds,'' \mpla{5}{90}{693}.}
\nref\ewtwog{E. Witten, ``On Quantum Gauge Theories in Two Dimensions,'' 
\cmp{141}{91}{153}; ``Two Dimensional Yang-Mills Theory Revisited,'' 
\jgp{9}{92}{303}, hep-th/9204083.}
\nref\duhe{V. Mathai and D. Quillen, ``Superconnections, Thom Classes, and 
Equivariant Differential Forms,'' {\it Topology} {\bf 25} (1986) 85\endli
M.F. Atiyah and L. Jeffrey, ``Topological Lagrangians and Cohomology,'' 
\jgp{7}{90}{119}.}
\nref\ampfine{A.M. Polyakov, ``Fine Structure of Strings,'' 
\npb{268}{86}{406}.}
\nref\rigidothers{H. Kleinert, ``The Membrane Properties of Condensing 
Strings,'' \plb{174}{86}{335}.}
\nref\ampbook{A.M. Polyakov, ``Gauge Fields and Strings'' (Harwood Academic 
Publishers, 1987).}
\nref\pisarski{R.D. Pisarski, ``Soluble Theory with Massive Ghosts,'' 
\prd{28}{83}{2547}; ``Smooth Strings at Large Dimension,'' \prd{38}{88}{578}; 
``Perturbative Stability of Smooth Strings,'' \prl{58}{87}{1300}; 
``Heavy and Smooth Strings in QCD,'' in: ``String Theory -- Quantum Cosmology 
and Quantum Gravity; Integrable and Conformal Invariant Theories,'' eds: H.J. 
de Vega and N. S\'anchez (World Scientific, 1987)\endli
E. Braaten, R.D. Pisarski and S.-M. Tse, ``Static Potential for Smooth 
Strings,'' \prl{58}{87}{93}.}
\nref\itoikub{C. Itoi and H. Kubota, ``BRST Quantization of the String 
Model with Extrinsic Curvature,'' \plb{202}{88}{381}; ``Gauge Invariance Based 
on the Extrinsic Geometry in the Rigid String,'' {\it Z. Phys.} {\bf C44} 
(1989) 337.}
\nref\oleseny{P. Olesen and S.-K. Yang, ``Static Potential in a String 
Model with Extrinsic Curvatures,'' \npb{283}{87}{73}.}
\nref\alonsoe{F. Alonso and D. Espriu, ``On the Fine Structure of Strings,'' 
\npb{283}{87}{393}.}
\nref\hightemp{J. Polchinski and Z. Yang, ``High Temperature Partition 
Function of the Rigid String,'' hep-th/9205043.}
\nref\wloops{S.-J. Rey and J. Yee, ``Macroscopic Strings as Heavy Quarks of 
Large N Gauge Theory and Anti-de~Sitter Supergravity,'' hep-th/9803001\endli
J.M. Maldacena, ``Wilson Loops in Lage N Field Theories,'' 
\prl{80}{98}{4859}, hep-th/9803002.}
\nref\ampcave{A.M. Polyakov, ``The Wall of the Cave,'' hep-th/9809057.}
\nref\parsour{G. Parisi and N. Sourlas, ``Random Magnetic Fields, 
Supersymmetry, and Negative Dimensions,'' \prl{43}{79}{744}; ``Self Avoiding 
Walk and Supersymmetry,'' \jpql{41}{80}{406}; ``Critical Behavior of Branched 
Polymers and the Lee-Yang Edge Singularity,'' \prl{46}{81}{871}.}
\nref\mckane{A.J. McKane, ``Reformulation of $n\rightarrow 0$ Models Using 
Anticommuting Scalar Fields,'' \pla{76}{80}{22}.}
\nref\susyphysica{N. Sourlas, ``Introduction to Supersymmetry in Condensed 
Matter Physics,'' \phsca{15}{85}{115}\endli
J.L. Cardy, ``Nonperturbative Aspects of Supersymmetry in Statistical 
Mechanics,'' \phsca{15}{85}{123}\endli
Y. Shapir, ``Supersymmetric Statistical Models on the Lattice,'' 
\phsca{15}{85}{129}.}
\nref\parstoch{G. Parisi and Y.-S. Wu, ``Perturbation Theory without 
Gauge Fixing,'' {\it Sci. Sinica} {\bf 24} (1981) 483\endli
G. Parisi and N. Sourlas, ``Supersymmetric Field Theories and Stochastic 
Differential Equations,'' \npb{206}{82}{321}.}
\nref\mitstoch{B. McClain, A. Niemi, C. Taylor and L.C.R. Wijewardhana, 
``Superspace, Dimensional Reduction, and Stochastic Quantization,'' 
\npb{217}{83}{430}.}
\nref\stochbook{M. Namiki, ``Stochastic Quantization,'' {\it L. N. Phys. 
Monographs} {\bf m9} (Springer Verlag, 1992).}
\nref\maststoch{J. Greensite and M.B. Halpern, ``Quenched Master Fields,'' 
\npb{211}{83}{343}\endli
M.R. Douglas, ``Stochastic Master Fields,'' \plb{344}{95}{117}, 
hep-th/9411025.}
\nref\polymers{P.G. de Gennes, ``Exponents for the Excluded Volume Problem 
as Derived by the Wilson Method,'' \pla{38}{72}{339}\endli
J. des Cloizeaux, ``Lagrangian Theory for a Self-Avoiding Random Chain,'' 
{\it Phys. Rev.} {\bf A10} (1974) 1665\endli
J. des Cloizeaux, ``The Lagrangian Theory of Polymer Solutions at 
Intermediate Concentrations,'' \jpq{36}{75}{281}.}
\nref\lubensky{T.C. Lubensky and J. Isaacson, ``Statistics of Lattice 
Animals and Dilute Branched Polymers,'' \pra{20}{79}{2130}.}
\nref\cloizbook{J. des Cloizeaux and G. Jannink, ``Les Polym\`eres en 
Solution: Leur Mod\'elisation et leur Structure'' (Les Editions de Physique, 
1987); English translation: ``Polymers in Solution: Their Modelling and 
Structure'' (Oxford University Press, 1990).}
\nref\pesando{I. Pesando and D. Cassi, ``Polymers and Topological Field 
Theory,'' \mplb{6}{92}{485}.}
\nref\nelsonpowers{P. Nelson and T. Powers, ``Renormalization of Chiral 
Couplings in Tilted Bilayer Membranes,'' \jpqII{3}{93}{1535}, 
cond-mat/9305014.}
\nref\clnp{W. Cai, T.C. Lubensky, P. Nelson and T. Powers, ``Measure Factors, 
Tension, and Correlations of Fluid Membranes,'' \jpqII{4}{94}{931}, 
cond-mat/9401020.}
\nref\jerusalem{``Statistical Mechanics of Membranes and Surfaces,'' 
Proceedings of the Jerusalem Winter School of Theoretical Physics, Vol.\ 5, 
eds: D. Nelson, T. Piran and S. Weinberg (World Scientific, 1988).}
\nref\david{F. David, ``Geometry and Field Theory of Random Surfaces and 
Membranes,'' in \jerusalem , p.\ 157\endli
F. David and S. Leibler, ``Vanishing Tension of Fluctuating Membranes,'' 
\jpqII{1}{91}{959}.}
\newsec{Introduction}

The AdS/CFT correspondence \refs{\maldacon - \gkp} relates quantum gravity in 
AdS spaces to superconformal field theories on the boundary of the AdS.  This 
intriguing relation between gravity and field theory has been much studied 
recently, and used as a new method for understanding strongly coupled gauge 
dynamics in various dimensions.  Among the cases with maximum supersymmetry 
are \maldacon\ the duality between $\CN=4$ $SU(N)$ super Yang-Mills theory in 
four dimensions and Type IIB theory on $\AdS_5\times\S^5$ with radius related 
to the 't~Hooft coupling via $(g_4^2N)^{1/4}$, and with $N$ units of RR flux 
through $\S^5$; and the duality between M-theory on $\AdS_7\times\S^4$ and the 
$(0,2)$ theory in six dimensions.  

In \refs{\ewadsholo,\ewtherm} this correspondence was further extended to 
supersymmetric theories at finite temperature and to non-supersymmetric 
theories.  In these cases, the dominant configuration on the AdS side 
is typically an AdS black hole with appropriate boundary conditions.  
Thus, non-supersymmetric $SU(N)$ Yang-Mills theory in four dimensions 
(pure $\QCD_4$) has been constructed by compactifying the six-dimensional 
$(0,2)$ theory to four dimensions on a torus with particular supersymmetry 
breaking periodicity conditions.  On the AdS side, the limit that correspond 
to pure $\QCD_4$ leaves one dimension asymptotic to $\S^1$ very small, and the 
M-theory background becomes a certain limit of the AdS Schwarzschild black 
hole in Type IIA string theory.  

These results provide some new support to the old suspicion that the master 
field of $\QCD_4$ may come in the form of a string theory.%
\foot{{}For some background on the idea of QCD strings, see e.g.\ 
\refs{\joereview,\ampconf}.}
However, the regime in which supergravity calculations are reliable 
corresponds to large values $\lambda\gg 1$ of the 't~Hooft coupling 
$\lambda=g_4^2N$, which translates on the gauge theory side to the strongly 
coupled gauge theory with 
a fixed UV cutoff $\Lambda$ set by the radius of the supersymmetry-breaking 
dimension \grosso .  The continuum limit of $\QCD_4$ lies in the regime where 
$\lambda$ is sent to zero as $\Lambda$ goes to infinity, 
\eqn\eeafads{\lambda\sim\,\frac{\beta_0}{\ln (\Lambda/\Lambda_{\QCD})},}
in accord with asymptotic freedom.  This limit is clearly difficult to 
understand on the string theory side, since it corresponds to Type IIA string 
theory in a rather singular limit of the corresponding AdS black hole 
background.  Moreover, it seems rather likely that the regimes with small and 
large $\lambda$ are separated by a phase transition \refs{\grosso - \miaoli}.  

Thus, the construction of the master field of pure $\QCD_4$ in the continuum 
limit relies, at least in this approach, on our ability to understand the 
$\lambda\rightarrow 0$ limit \eeafads\ on the string theory side of the 
correspondence.  The AdS construction indicates that QCD glueballs should 
correspond on the Type IIA string theory side to supergravitons \grosso\ in 
the background of the AdS black hole.  In the $\lambda\rightarrow 0$ limit of 
continuum $\QCD_4$, we expect that the lowest closed string state should 
develop a non-zero mass gap of order $\Lambda_{\QCD}$.  Any direct 
understanding of such ``dressing'' of the supergraviton seems very difficult 
in the elementary string theory approach, in particular if we try to 
extrapolate from the $\lambda\gg 1$ regime of small curvature.  

The AdS/CFT correspondence may provide a microscopic theory of the QCD master 
field in principle, in terms of the corresponding limit of the Type IIA string 
background.  One can certainly hope that a detailed understanding of the 
Type IIA string theory sigma model in the RR background and in the singular, 
large curvature limit will give satisfactory answers to some of these puzzling 
questions.  However, to 
understand the continuum limit of $\QCD_4$ using the AdS construction, we do 
not need to understand Type IIA string theory on this class of backgrounds; 
this class contains non-universal information which will be irrelevant in the 
continuum $\QCD_4$ limit.  {}For example, in the continuum limit we expect 
various states of the full elementary string theory -- in particular, the 
$SO(6)$ non-singlet Kaluza-Klein modes on $\S^5$ -- to decouple from the part 
of the spectrum corresponding to QCD.  

If we are only interested in the universal properties of continuum QCD, 
and given the obvious difficulty with understanding the full Type IIA string 
theory on this class of backgrounds, it may be natural to search 
for alternatives, in the form of an effective string theory that would 
directly describe the rather difficult singular limit of Type IIA string 
theory.  The purpose of this paper is to initiate some steps in such an 
effective string theory approach.  

In Section~2 we present some general discussion of scales in the hypothetical 
QCD string theory, and use the AdS/CFT relation to derive a ``UV-UV 
correspondence'' between the string world-sheet and the space-time of QCD.  

In Section~3, we discuss the case of pure $\QCD_2$ at large-$N$.  
In this particular case, a string theory whose path integral reproduces the 
large-$N$ expansion of $\QCD_2$ is known \refs{\trst - \srst}.  
This theory was originally constructed as a topological rigid string theory, 
but we will see later in the paper that it is more natural to think of the 
theory of \refs{\trst,\srst}  as a rigid string theory with world-sheet 
Parisi-Sourlas supersymmetry.  We review some of the properties of this 
two-dimensional rigid string theory as a theory of $\QCD_2$ strings in 
Section~3.1.  In Section~3.2, using the AdS/CFT technology, we construct pure 
$\QCD_2$ by compactifying $\CN=4$ super Yang-Mills theory from four 
dimensions.  In the corresponding limit of pure $\QCD_2$ at large $N$ and 
fixed but small 't~Hooft coupling, the theory should be described by Type IIB 
string theory on an asymptotically AdS background in 
a rather singular limit.  In Section~3.3 we compare the two approaches, and 
conjecture that Type IIB string theory in the singular limit of the AdS 
background is equivalent to the supersymmetric rigid string of Section~3.1.  

Given this relation between the supersymmetric rigid string theory on one 
hand, and the Type IIB string theory in the singular background corresponding 
to $\QCD_2$ on the other, we study possible extensions of the supersymmetric 
rigid string theory to higher dimensions.  In Section~4, we find a very 
natural generalization of the supersymmetric rigid string of Section~3.1 to 
dimensions greater than two, in the form of a rigid string theory with local 
Parisi-Sourlas supersymmetry. Symmetries of this theory do not allow any 
tension term in the Lagrangian, and the theory has only one marginal 
coupling -- the string rigidity $\alpha_0$.  We calculate the one-loop beta 
function for $\alpha_0$, and find that the theory is asymptotically free in 
dimensions greater than two.  We argue that due to dimensional transmutation 
on the world-sheet, a non-zero string tension is dynamically generated at 
large distances, and supersymmetry is broken.  In contrast, the 
two-dimensional theory is supersymmetric at all scales, and the Parisi-Sourlas 
supersymmetry effectively plays the role of topological BRST invariance.  In 
dimensions higher than two, the string has physical oscillations due to 
non-zero tension at large distances, but becomes asymptotically topological 
due to the restoration of supersymmetry in the UV.  

Even though the main motivation for our study of the supersymmetric rigid 
string in dimensions higher than two originates largely from the arguments 
presented earlier in Sections~1 -- 3 of the paper, Section~4 can be read 
essentially independently of the rest of the paper.   

\newsec{Correspondence of Scales in QCD String Theory}

The AdS/CFT correspondence relates a theory with gravity in the bulk to a 
gauge theory at the boundary.  As was pointed out in \lsew\ (and further 
refined in \apjp ), the precise holographic behavior of the bulk gravity 
theory is possible because of a UV-IR correspondence between the scale in the 
boundary theory and the scale in the elementary string theory (or M-theory) in 
the bulk.   In the AdS/CFT correspondence, infrared effects in the bulk theory 
map to ultraviolet effects in the boundary theory \lsew .  

Elementary string theory itself is known to exhibit another UV-IR 
correspondence, which relates the ultraviolet regime on the string 
world-sheet to the ultraviolet effects in space-time.  One way to 
see this heuristically is in the logarithmic behavior of the world-sheet 
propagator, $\left\langle X(0)X(\sigma )\right\rangle\sim\log\sigma$.  
This UV-IR correspondence between the world-sheet of the elementary 
string and the space-time in which the string propagates should also hold in 
the Type IIB theory on $\AdS_5\times\S^5$, at least for large $g_sN$.  
In combination with the UV-IR correspondence of \lsew , this UV-IR 
correspondence allows us to derive a direct correspondence of scales 
between the world-sheet of the string and the space-time of QCD.  Thus, we 
obtain the following {\bf UV-UV correspondence}:

\noindent{\it Ultraviolet effects on the string world-sheet are mapped to 
ultraviolet effects in QCD.}

This behavior is not surprising, and its validity in the hypothetical 
QCD string theory has indeed been anticipated.  Here, however, we have 
{\it derived\/} this anticipated correspondence of scales in QCD string 
theory from the microscopic theory, using a very simple chain of arguments 
based on the AdS/CFT correspondence.  

Having argued that the ultraviolet regime on the string world-sheet should 
map to the ultraviolet effects in QCD, it seems somewhat awkward that 
the AdS construction leads to a microscopic description of gauge theory 
in terms of elementary string theory, in which the anticipated UV-UV 
correspondence is rather obscure.  It would seem natural to look for 
another string theory description, in which the UV-UV correspondence between 
the string world-sheet and the QCD space-time would be more manifest.  

One such theory was proposed a few years ago, in the form of a particular 
supersymmetrization of the rigid string.  This theory was tested in the case 
of pure-glue QCD in two dimensions, and was found to precisely reproduce the 
exact results of its large-$N$ expansion previously obtained in \grta .  

\newsec{Strings and $\QCD_2$}

Two-dimensional pure QCD is an exactly solvable theory \refs{\migrus,\ewtwog}
on any Riemann surface%
\foot{Throughout this paper we will work in Euclidean signature.}
and for any (compact) gauge group.  In the case of $SU(N)$, its large-$N$ 
expansion was analyzed in \grta , and it was shown that the partition 
function as well as correlation functions of Wilson loops can be expressed as 
an infinite sum over maps from auxiliary world-sheets to the $1+1$ dimensional 
space-time.  Schematically, the partition function for $\QCD_2$ on a surface 
of genus $G$ can be written in the large-$N$ expansion as \grta
\eqn\eeqcdtwopart{\CZ_G=\sum_{g}\frac{1}{N^{2g-2}}\sum_{n,\tilde n}e^{-\lambda 
A(n+\tilde n)/2}\sum_{i=0}^{t}\left(\frac{\lambda A}{2}\right)^{i}
\omega_{i,g,G,i,n,\tilde n}.}
Here $\lambda=g_2^2N$ is the two-dimensional 't~Hooft coupling, $A$ is the 
area of the space-time, and $1/N$ is the string coupling constant. 
The string world-sheet covers the space-time $n$ times in the orientation 
preserving sector, and $\tilde n$ times in the orientation reversing sector; 
$n+\tilde n>0$.  The upper limit $t$ in the last summation depends on $g$, 
$G$, $n$ and $\tilde n$.  $\omega$ are numerical coefficients whose precise 
form can be found in \grta . 

In \refs{\trst,\srst}, this exact large-$N$ expansion was found to correspond 
to the path integral of a new kind of rigid string theory.  An alternative 
approach to this problem was presented in \cmr .  The results of \refs{\trst 
- \srst} give us the first known example of a complete world-sheet 
Lagrangian description of QCD string theory.  

\subsec{Large-$N$ Expansion in $\QCD_2$ as a Rigid String Theory}

In this section we review some of the highlights of the world-sheet 
representation of $\QCD_2$ strings, in the framework of \refs{\trst,\srst}.  
Our discussion will be relatively brief; the reader can find details in 
\srst .  

One starts with the crucial observation \trst\ that in the large-$N$ expansion 
\eeqcdtwopart , only minimal-area surfaces contribute.  The world-sheet theory 
should therefore be expected to localize to minimal-area maps.  Consider maps 
from a given world-sheet $\Sigma$ with coordinates $\sigma^a$ to a fixed 
space-time manifold with coordinates $x^\mu$ and fixed 
metric $g_{\mu\nu}$ (not necessarily flat).  The minimal-area condition 
\eqn\eeminarea{\Delta x^\mu=0,}
where $\Delta$ is the Laplace operator with respect to the induced 
world-sheet metric $h_{ab}=\p_ax\cdot\p_bx$, is almost empty in two target 
dimensions, but not quite.  Its main purpose is to forbid maps with folds of 
non-zero length.  The only moduli of minimal-area maps are the locations of 
various singularities of the maps (such as branched points).  

There is a well-known technique for constructing path integrals that 
localize to solutions of a given equation \duhe .  Following this procedure 
in the case of the minimal-area condition \eeminarea , one derives a certain 
supersymmetric world-sheet theory \refs{\trst,\srst}.  This theory is 
described by the following world-sheet Lagrangian 
\eqn\eetrstlag{\eqalign{\CL&=\frac{1}{\alpha_0}\int_\Sigma d^2\sigma\,\sqrt 
h\left\{\frac{1}{2}\Delta x\cdot\Delta x-\chi\cdot\Delta\psi-
R_{\mu\nu\sigma\rho}\chi^\mu\psi^\nu\chi^\sigma\psi^\rho\right.\cr
&\qquad\qquad\left.{}+\left(h^{ab}h^{cd}-h^{ac}h^{bd}-h^{ad}h^{bc}\right)
\p_a\chi\cdot\p_bx\;\p_c\psi\cdot\p_dx+\ldots\phantom{\frac{1}{2}}\right\},
\cr}}
where the fermionic fields $\psi^\mu$ and $\chi^\mu$ are world-sheet scalars, 
the $\cdot$ denotes the inner product with respect to $g_{\mu\nu}$.  We 
have left out terms of higher order in fields that are not relevant for 
our discussion here; the full Lagrangian in curved space-time can be found 
in \srst .  

In addition to world-sheet diffeomorphism symmetry, the Lagrangian \eetrstlag\ 
is invariant under two mutually anticommuting nilpotent supersymmetries, 
\eqn\eesusyalg{\eqalign{[Q,x^\mu]&=\psi^\mu,\cr
\{Q,\psi^\mu\}&=0,\cr
\{Q,\chi^\mu\}&=\Delta x^\mu+\ldots,\cr}
\qquad
\eqalign{[\bar Q,x^\mu]&=\chi^\mu,\cr
\{\bar Q,\psi^\mu\}&=-\Delta x^\mu+\ldots,\cr
\{\bar Q,\chi^\mu\}&=0.\cr}}
In the topological interpretation of the theory, $Q$ is of course the BRST 
charge, while $\bar Q$ would be an ``accidental'' anti-BRST symmetry.  

The kinetic term $\sim (\Delta x)^2$ in \eetrstlag\ is the famous extrinsic 
curvature term that appears in the rigid string Lagrangian \ampfine\ (see 
also \refs{\rigidothers - \hightemp} for more on the bosonic rigid string).  
The dimensionless coupling constant $\alpha_0$ is usually called the string 
rigidity.  It is rather striking that we have {\it derived}, from the sole 
requirement of localization to minimal-area maps, a certain supersymmetrized 
version of the rigid string Lagrangian!  

The on-shell symmetries \eesusyalg\ can easily be promoted into off-shell 
symmetries of the theory, by reintroducing an auxiliary field $B^\mu$, 
and rewriting the Lagrangian (for simplicity in flat space-time) as 
\eqn\eetwo{\eqalign{\CL&=\frac{1}{\alpha_0}\int_\Sigma d^2\sigma\sqrt h\left\{
h^{ab}\p_aB\cdot\p_bx+h^{ab}\p_a\psi\cdot\p_b\chi\right.\cr
&\qquad\qquad\left.{}+\left(h^{ab}h^{cd}-h^{ac}h^{bd}-h^{ad}h^{bc}\right)
\p_a\chi\cdot\p_bx\;\p_c\psi\cdot\p_dx+B^2{\rm \ terms}\right\}.\cr}}
The off-shell supersymmetry algebra is identical with \eesusyalg , with 
the $\Delta x^\mu+\ldots$ terms replaced by $B^\mu$.  Of course, $B^\mu$ 
satisfies $[Q,B^\mu ]=[\bar Q,B^\mu ]=0$.  

The ordinary bosonic rigid string Lagrangian also contains a tension term 
\ampfine .  In our supersymmetric rigid string theory, no tension term is 
allowed by the off-shell supersymmetry of \eetwo .  

Physical states are defined as the cohomology classes of the BRST operator 
$Q$, with an extra condition of gauge invariance imposed by world-sheet 
diffeomorphism symmetry.  These conditions on physical states leave precisely 
one bosonic state in each non-trivial winding sector of the 
closed string, in accord with the known spectrum of states in $\QCD_2$ 
string theory.  

Almost by construction (or more precisely, by world-sheet supersymmetry 
\eesusyalg ), the path integral localizes to the moduli space of minimal-area 
maps.  One can calculate the path integral in the semi-classical, $\alpha_0
\rightarrow 0$ limit of the world-sheet theory.  The partition function 
will not depend on $\alpha_0$.  It is not difficult to see that the path 
integral in fact calculates the Euler number of the moduli spaces of 
minimal-area maps \refs{\trst,\srst}.  Indeed, recall that the zero modes of 
$\psi^\mu$ behave as one-forms on the moduli spaces.  Upon integration over 
the zero modes of $\chi^\mu$ and $\psi^\mu$, the curvature-dependent term in 
\eetrstlag\ will give the Euler number density on the moduli space.  

It was conjectured in \trst\ that the coefficients in the large-$N$ expansion 
\eeqcdtwopart\ at $\lambda A=0$ are precisely the Euler numbers of the moduli 
spaces of minimal-area maps, and that the path integral in the topological 
rigid string theory reproduces the exact results of the large-$N$ expansion 
of $\QCD_2$.   The Euler numbers of the relevant moduli spaces were first 
obtained in \cmr , and the conjecture was later proven in \srst . 

The case of non-zero $\lambda A$ is a little more involved, but it can also be 
analyzed in the framework of topological string theory.  The structure of the 
large-$N$ expansion \eeqcdtwopart\ indicates that the corresponding string 
theory is still localized to minimal-area maps, and suggests that the 
polynomial area dependence arises from measuring the volume of various 
components of the moduli space.  This can be represented in the topological 
rigid string as follows.  Due to localization to moduli spaces, one can 
calculate correlation functions of observables that correspond to universal 
cohomology classes on the moduli spaces; a particular linear combination of 
these universal observables then corresponds to the tension term.  The 
exponential dependence on $A$ in \eeqcdtwopart\ is reproduced from the 
evaluation of $\int_\Sigma d^2\sigma\sqrt{h}$ on the minimal-area map.  
{}For details on the perturbative treatment of the $\lambda A\neq 0$ case in 
the topological rigid string theory, the reader is 
again referred to \srst .  We will have more to say about the area dependence 
in $\QCD_2$ string theory in Section~4.5.

\subsec{AdS Construction for $\QCD_2$}

Using the AdS/CFT correspondence, and following the construction of 
$\QCD_4$ in \ewtherm , we could in principle construct $\QCD_2$ in several 
ways.  One option is to start with $N$ M2-branes at finite temperature, i.e.,  
with the superconformal theory in three dimensions at the strongly-coupled 
fixed point with $SO(8)$ R-symmetry, compactified on an $\S^1$  with 
supersymmetry-breaking periodicity conditions.  According to the 
AdS/CFT correspondence, this theory would correspond to the Schwarzschild 
black hole in the $\AdS_4$ space-time (times the internal $\S^7$).  

Here we will start instead with $\CN=4$ super Yang-Mills in four dimensions, 
and compactify this theory, first to three dimensions on $\S^1$ with radius 
$R_1$ and supersymmetry-preserving conditions, and subsequently to two 
dimensions on $\S^1$ with radius $R_2$ and supersymmetry-breaking conditions.  
The first step leads to dimensionally reduced Yang-Mills theory with sixteen 
supercharges, at distances much larger than $R_1$.  Keeping $R_2\gg R_1$, 
and looking at the system at distances much larger than $R_2$ in the remaining 
two non-compact dimensions, the second step should give masses of order 
$1/R_2$ to the fermions, and of order $g_4^2N/R_2$ to the scalars.  

Since we would like to compare the theory to $\QCD_2$ on a two-torus of finite 
area $A$, we will further compactify the two remaining dimensions on a 
two-torus with both radii of order $L=\sqrt A$ (say on a rectangular torus 
with both radii equal to $L$).  The radius $L$ of the two-torus sets the scale 
in $\QCD_2$.  

In order to isolate $\QCD_2$ we want to keep $L$ fixed while sending 
$R_1\ll R_2\rightarrow 0$.  The $\QCD_2$ coupling $g_2$ is related 
to the four-dimensional super Yang-Mills coupling via $(2\pi^2)g_2^2=
g_4^2/R_1R_2$.  The two-dimensional 't~Hooft coupling $\lambda=g_2^2N$  is 
then given in terms of the four-dimensional quantities by 
\eqn\eetensten{\lambda=\frac{g_4^2N}{(2\pi)^2R_1R_2}.}

We are interested in the physics at scales of order $L$, in the 
asymptotic large-$N$ expansion, and at zero QCD string tension $\lambda$.  
Thus, we want to take the limit where $L$ is fixed, $R_1,R_2\rightarrow 0$, 
and $N\gg 1$ but finite.  

In order to make contact with the supersymmetric rigid string description of 
$\QCD_2$, we are interested in the limit of $\lambda\rightarrow 0$, in which 
as we have seen in the previous section the string path integral 
representation is most natural.  $\QCD_2$ on a torus is not completely trivial 
even in the tensionless limit $\lambda A=0$, as one can study correlation 
functions of Wilson lines.  However, it is still convenient to keep 
$\lambda A$ small but non-zero, as a regulator in the sum over world-sheets 
of various winding numbers $n$, $\tilde n$ in \eeqcdtwopart .  

Thus, we need $\lambda$ to be small in units set 
by $L$, which means that $g_4$ will have to scale with $R_1$ and $R_2$ 
as $g_4^2\sim R_1R_2$.  In order to keep $\lambda$ small in the infinite $N$ 
limit, $g_4$ will have to scale with $N$ as well: $g_4^2\sim R_1R_2/N$.  
The AdS/CFT relation identifies $(2\pi)^{-1}g_4^2$ with the Type IIB string 
coupling constant $g_s$, and we see that in the limit of our interest the 
Type IIB string theory is indeed very weakly coupled.  

As $R_1$ and $R_2$ go to zero at fixed small $\lambda$, the masses of 
fermions and scalars go to inifinity, and the states should indeed decouple.  
The effective bare coupling at the compactification radius is 
$g_2^2N(2\pi)^2R_1R_2=g_4^2N$, which is also small in our limit.  

The five-dimensional metric of the corresponding Type IIB string theory is 
\eqn\eefivemet{ds^2=\left(\frac{\rho^2}{b^2}-\frac{b^2}{\rho^2}\right)d\tau^2
+\frac{d\rho^2}{\left(\frac{\rho^2}{b^2}-\frac{b^2}{\rho^2}\right)}+\rho^2
\sum_{i=1}^3dx_i^2.}
The period of $\tau$ is $\beta_1=\pi b$, $b$ is related to the Type IIB 
string coupling $g_s$, the string scale $\ell_s$ and the RR flux $N$ by 
$2b=\ell_s(4\pi g_sN)^{1/4}$, and 
we have temporarily suppressed the $\S^5$ part of the full Type IIB string 
theory.  

It will be convenient to rescale $\tau$ so that the rescaled coordinate 
$\theta=\ell_s^{-1}(4\pi g_sN)^{-1/4}\tau$ has period $2\pi$.  We also 
introduce a rescaled coordinate $r=2\rho/b$, and the metric becomes
\eqn\eefivemett{ds^2/\ell_s^2=
(4\pi g_sN)^{1/2}\left({r^2}-\frac{1}{r^2}\right)d\theta^2
+(4\pi g_sN)^{1/2}\frac{dr^2}{\left(r^2-\frac{1}{r^2}\right)}+
(4\pi g_sN)^{1/2}r^2\sum_{i=1}^3dx_i^2.}

So far we have compactified only one dimension in the boundary field theory, 
on a circle with supersymmetry-breaking boundary conditions.  We also want 
to compactify the remaining Killing dimensions $x_i$ on a three-torus, with 
one dimension (say $x_3\equiv y$) smaller than $\tau$, and the two remaining 
$x_{1,2}$ much larger than $\tau$.  The radius of $\theta$ at infinity can 
of course be arbitrarily changed by rescaling $r$ and defining the 
boundary metric as $r^{-2}ds^2$ in the $r\rightarrow 0$ limit.  

We obtain the correct compactification by requiring that the radius of $y$ 
at infinity be $R_2/R_1$ times the radius of $\theta$ at infinity (and 
similarly for $x_i$).  It will again be convenient to rescale $y$ and $x_i$ 
such that the rescaled variables $Y$ and $X_i$ have period $2\pi$.  In 
terms of these variables, the fully compactified metric is now 
\eqn\eefivemetcomp{\eqalign{ds^2/\ell_s^2&=
(4\pi g_sN)^{1/2}\left({r^2}-\frac{1}{r^2}\right)d\theta^2
+(4\pi g_sN)^{1/2}\frac{dr^2}{\left(r^2-\frac{1}{r^2}\right)}\cr
&\qquad\qquad\qquad{}+(4\pi g_sN)^{1/2}r^2\left(\left(\frac{R_2}{R_1}
\right)^2dY^2+\left(\frac{L}{R_1}\right)^2\sum_{i=1}^2dX_i^2\right).\cr}}

Now we are ready to study this metric in the limit that should isolate only 
the $QCD_2$ degrees of freedom.  We have argued that in this limit, as 
$R_1$ and $R_2$ go to zero, the Type IIB string coupling is given by 
\eqn\eegstr{g_s=2\pi \frac{\lambda R_1R_2}{N}.}
Using this formula, one can see that all dependence on $N$ disappears from 
the metric, which now becomes 
\eqn\eefivemetcnon{\eqalign{\frac{ds^2}{2(2\pi)^2\ell_s^2}=
\lambda^{1/2}(R_1R_2)^{1/2}\left({r^2}-\frac{1}{r^2}\right)&d\theta^2
+\lambda^{1/2}(R_1R_2)^{1/2}\frac{dr^2}{\left(r^2-
\frac{1}{r^2}\right)}\cr
&{}+
\lambda^{1/2}r^2\left(\frac{R_2^{5/2}}{R_1^{3/2}}dY^2+
\frac{L^2R_2^{1/2}}{R_1^{3/2}}\sum_{i=1}^2dX_i^2\right).\cr}}
In this picture, $N$ appears only in the expression for the string coupling 
\eegstr , leading to the $g_s\sim 1/N$ relation anticipated in QCD string 
theory.  

We will take the ``continuum $\QCD_2$'' limit by sending $R_2\equiv R$ to 
zero while keeping $R_1/R_2\ll 1$ fixed.  In this limit, the metric scales 
with $R\rightarrow 0$ as follows, 
\eqn\eefivemetcnon{\frac{ds^2}{2(2\pi)^2\ell_s^2}=
\lambda^{1/2}R\left({r^2}-\frac{1}{r^2}\right)d\theta^2
+\lambda^{1/2}R\frac{dr^2}{\left(r^2-\frac{1}{r^2}\right)}+
\lambda^{1/2}Rr^2dY^2+\lambda^{1/2}\frac{L^2}{R}r^2
\sum_{i=1}^2dX_i^2.}
As $R$ goes to zero, only the two dimensions corresponding to the space-time 
of $\QCD_2$ are large.  All the remaining dimensions (including the $\S^5$ 
that is implicit in our formulas) are suppressed by a positive power of 
$R$.  Of course, we still have $1\leq r\leq\infty$.  

This little construction of $\QCD_2$ from the AdS/CFT correspondence 
illustrates one of the main points that we want to make: decoupled QCD in the 
continuum limit corresponds to a rather singular limit on the AdS side 
of the AdS/CFT correspondence, a limit in which supergravity is 
manifestly invalid.  This conclusion is fairly independent of the 
space-time dimension, and applies equally well to the construction of 
continuum $\QCD_4$ from the six-dimensional $(0,2)$ theory compactified on 
a two-torus \ewtherm , even though our reason for sending $\lambda\rightarrow 
0$ in that case is different \grosso .  

As we have argued in the Introduction, continuum QCD might be described by 
elementary string theory only in this singular limit, which is manifestly 
hard to understand.  We may therefore want to describe it directly by another 
string theory that is more suitable for the description of the surviving 
degrees of freedom in that singular limit of Type IIB string theory.  

\subsec{Comparison}

In the case of $\QCD_2$, we already have an answer for the world-sheet theory 
of QCD strings, in the form of the supersymmetric rigid string reviewed in 
Section~3.1.  In conjunction with the AdS construction of $\QCD_2$ presented 
in Section~3.2, this leads to the following conjecture:  

{\it The two-dimensional supersymmetric rigid string theory  
of \refs{\trst,\srst} is dual to the elementary Type IIB string 
theory compactified on the black-hole geometry asymptotic to 
$\AdS_5\times\S^5$ in the singular limit described in Section~3.2.}

In the sense of this conjecture, the supersymmetric rigid string theory 
of $\QCD_2$ represents a ``boundary condition'' on any candidate QCD string 
theory that we may want to derive from Type II string theory on the singular 
backgrounds that correspond to the continuum limit of QCD in higher 
dimensions.  

The supersymmetric rigid string theory belongs to the category of string 
theories whose kinetic term is given by the extrinsic curvature term 
$(\Delta x^\mu)^2$.  Such theories exhibit a UV-UV correspondence between 
the world-sheet of the string and the space-time in which the string 
propagates.  One can see this fact simply from the structure of the 
Lagrangian for $x^\mu$ at small $\alpha_0$.  Due to the higer-order form of 
the kinetic term, $x^\mu$ has the canonical dimension of length on the 
world-sheet.  This should again be contrasted with elementary string theory, 
which exhibits the UV-IR correspondence.  

The supersymmetric rigid string was constructed from the requirement of 
localization to moduli spaces of minimal-area surfaces.  This localization 
has an intriguing analogy in the AdS construction of $\QCD_2$.  In the AdS 
construction, the elementary string tension is related to the 
effective string tension, as seen in the Wilson loop average, by a factor 
of $b^2$ \refs{\wloops,\grosso}.  As we send $b$ to zero and keep the 
effective string tension (small but) finite in the units set by $L$, the 
elementary string tension goes to infinity.  Thus, the elementary string is 
classical on the world-sheet, and its world-sheet path integral should be 
dominated by its classical solutions.  

In the path integral of the supersymmetric rigid string, one of the main 
subtleties comes from the fact that any minimal-area map between 
two-dimensional surfaces in a generic homotopy class exhibits singularities 
where the induced metric is degenerate.  Due to this singular nature of 
generic minimal-area maps, a more precise definition of the class of surfaces 
that contribute to the path integral is needed.  The correct choice is the 
class of so-called stratified surfaces \srst .  A related phenomenon has been 
observed in \grosso , where the calculation of Wilson lines in the 
supergravity regime leads to collapsed minimal surfaces with two disk 
components connected by an infinitesimally thin tube, which is -- at scales 
large compared to the string scale -- a typical stratified surface.  Using the 
conjectured relation between the Type IIB string theory and the supersymmetric 
rigid string, the full Type IIB string theory on the background slightly away 
from its singular limit should provide a covariant regularization of this 
problem of singular world-sheets in the path integral of the supersymmetric 
rigid string.    

It would be extremely interesting to see whether the supersymmetric rigid 
string theory can be derived from the Type IIB string theory sigma model 
in the corresponding singular limit, once a precise formulation of the 
latter becomes available at least in the limit of large 't~Hooft coupling.  

\newsec{Rigid String Theory with Parisi-Sourlas Supersymmetry}

Given that the supersymmetric rigid string approach has been successful in 
describing $\QCD_2$ strings, one is naturally interested in possible 
generalizations of the theory to dimensions greater than two.  The purpose of 
this Section is to show that at least one natural extension of the 
supersymmetric rigid string theory does indeed exist.  This rigid string 
theory is characterized by local Parisi-Sourlas supersymmetry on the 
world-sheet.  We will see that in dimensions greater than two, this theory 
exhibits very interesting dynamics, involving asymptotic freedom and 
dimensional transmutation.  Consequently, the theory is effectively 
topological only in two space-time dimensions, where it reproduces the 
$\QCD_2$ string theory reviewed in Section~3.1.  

Some ideas presented in this Section have also been advocated independently 
by Polyakov (\refs{\ampconf,\ampcave} and unpublished).  

\subsec{A Superspace Formulation}

Our starting point will be the supersymmetric rigid string Lagrangian \eetwo , 
in flat space-time $\R^D$. 

The rigid string Lagrangian \eetwo\ can be succinctly rewritten in a 
superspace form if we assemble all the fields in the Lagrangian into $D$ 
superfields,
\eqn\eethree{X^\mu(\sigma^a,\theta,\bar\theta)=x^\mu+\theta\psi^\mu+
\bar\theta\chi^\mu +\theta\bar\theta B^\mu.}
Here $\theta,\bar\theta$ (which we will sometimes collectively denote by 
$\theta^\alpha$) are two fermionic coordinates that transform as scalars 
with respect to ordinary world-sheet diffeomorphisms.%
\foot{i.e., they form a trivial two-dimensional bundle over $\Sigma$, with 
odd fiber.  Since this trivial bundle is unique, we will not distinguish in 
our notation between $\Sigma$ and its super-extension.}

In this notation, we can write the Lagrangian of the topological rigid string 
as a sum of the following two terms, $\CL_1$ and $\CL_2$.  
The first term, $\CL_1$, looks like the ordinary bosonic Nambu-Goto 
Lagrangian, with the target coordinates replaced by the superfields $X^\mu$, 
and integrated over the whole super world-sheet:  
\eqn\eefour{\CL_1=\frac{1}{\alpha_0}\int_\Sigma d^2\sigma d^2\theta\sqrt{\det 
H_{ab}}.}
Here $d^2\theta\equiv d\bar\theta d\theta$, and $H_{ab}\equiv H_{ab}
(\sigma^a,\theta^\alpha)$ is the induced metric on the world-sheet:
\eqn\eefive{H^{ab}\equiv\p_a X^\mu g_{\mu\nu}\p_b X^\nu\equiv\p_aX\cdot\p_bX.}

This Lagrangian is invariant under a supersymmetric extension of the 
world-sheet diffeomorphism group.  It is not, however, invariant under all 
superdiffeomorphisms of the super world-sheet.  The group of symmetries of 
the Lagrangian consists of arbitrary reparametrizations of $\sigma^a$,
\eqn\eesix{\tilde\sigma^a=\tilde\sigma^a(\sigma^b,\theta,\bar\theta),\qquad
\tilde\theta^\alpha=\theta^\alpha,}
and all ($\sigma^a$-independent) diffeomorphisms of $\theta^\alpha$, 
\eqn\eeseven{\tilde\theta^\alpha=\tilde\theta^\alpha(\theta^\beta).}
Hence, although the new bosonic coordinates $\tilde\sigma^a$ can be arbitrary 
regular functions of both $\sigma^a$ and $\theta^\alpha$, the new fermionic 
coordinates $\tilde\theta^\alpha$ are $\sigma^a$-independent functions 
of $\theta^\beta$ only.  This ``semirigid supersymmetry'' \srst\ arises quite 
naturally in topological field theories with diffeomorphism invariance, as 
it is the smallest group that combines global BRST symmetry with 
diffeomorphisms. 

So far we have found the superspace expression for only a part of the 
Lagrangian of the topological rigid string.  The requirement of invariance 
under semirigid world-sheet supersymmetry gives us one more building block 
for the construction of Lagrangians.  It is easy to see that 
\eqn\eeeight{S\equiv\p_\theta X\cdot\p_{\bar\theta}X-\p_\theta X\cdot\p_a X\;
H^{ab}\;\p_b X\cdot\p_{\bar\theta}X}
is invariant under the semirigid supersymmetry \eesix , \eeseven .  Probably 
the easiest way to see this invariance is to introduce the induced supermetric 
on the super world-sheet (here we introduce the superspace index on the 
world-sheet, $A\equiv (a,\alpha)$, with $a=1,2,\ \alpha=1,2$):
\eqn\eenine{H_{AB}=(-1)^B\p_AX\cdot\p_BX,}
and notice that $S$ is the inverse of the $\theta\bar\theta$ component of 
$H^{AB}$, where $H^{AB}$ is the inverse of the induced supermetric, 
\eqn\eeten{S=(H^{\theta\bar\theta})^{-1}.}
Invoking the transformation properties of the supermetric, $S$ is invariant 
under semirigid superdiffeomorphisms.  We can use this invariant to write 
\eqn\eeeleven{\CL_2=\frac{1}{\alpha_0}\int_\Sigma d^2\sigma d^2\theta
\sqrt{\det H_{ab}}S.}
While the $\theta\bar\theta$ component of $S$ produces the $B^2$ term of the 
topological rigid string Lagrangian \eetwo .  

Thus, we have succeeded in rewriting the theory of Section~3.1 as a rigid 
string theory with semirigid supersymmetry.  The semirigid supersymmetry 
algebra contains two scalar mutually anticommuting supercharges $Q$ and 
$\bar Q$ of \eesusyalg\ which act on the superfields \eethree\ by 
\eqn\eepsalg{Q=\frac{\p}{\p\theta},\qquad\bar Q=\frac{\p}{\p\bar\theta}.}
This type of supersymmetry, known in the literature as the Parisi-Sourlas 
supersymmetry \refs{\parsour - \susyphysica}, is a very universal symmetry of 
many physical systems.  It has a remarkably broad range of applications, in 
the physics of polymers, quenched disorder, systems in random external fields, 
spin glasses, and stochastic quantization, to name just a few.  (See e.g.\ 
\refs{\parsour - \pesando} for some references on this subject.)  In 
particular, the way in which our complicated higher-order rigid string 
Lagrangian \eetrstlag\ can be 
rewritten in the Parisi-Sourlas superspace in terms of a supersymmetrized 
Nambu-Goto Lagrangian \eefive\ (and \eeeleven ) is also very reminiscent of 
similar constructions in theories with Parisi-Sourlas supersymmetry.  In the 
case of our rigid string, Parisi-Sourlas supersymmetry has been combined with 
diffeomorphism invariance, into the semirigid supersymmetry of the theory.  

Having recognized that our theory is in fact a string theory with world-sheet 
Parisi-Sourlas supersymmetry, we now want to write down the most general 
world-sheet Lagrangian that respects the symmetry (together with world-sheet 
diffeomorphisms and space-time Poincar\'e invariance) and contains only 
marginal or relevant terms.  

The structure of the Lagrangian \eefour\ near the gaussian fixed point at 
$\alpha_0=0$ determines canonical world-sheet dimensions as follows.  
$d\sigma^a$ has the dimension of world-sheet length, as does $\theta^\alpha$.  
Since $d\theta\sim \p\theta$, the super-volume element $d^2\sigma d^2\theta$ 
is dimensionless.  All superfields $X^\mu$ have the dimension of world-sheet 
length, and $\alpha_0$ is indeed dimensionless. 

With this counting, $S$ of \eeeight\ is of canonical dimension zero, and the 
general theory is described by the following Lagrangian: 
\eqn\eetwelve{\CL=\int_\Sigma d^2\sigma d^2\theta\sqrt{\det H_{ab}}\left(
\frac{1}{\alpha_0}+\sum g_{i0}S^i\right),}
where the sum goes over all positive integer powers of $S$, and $g_{i0}$ 
are independent coupling constants.  There are no relevant terms that would 
respect the symmetries.  

At first glance, it might seem that in \eetwelve\ we have a theory with an 
infinite number of canonically marginal coupling constants $g_{i0}$; however, 
the number of independent ones is finite since $S^n$ vanishes for $n$ high 
enough.  The analysis of this Lagrangian at the quantum level and the 
identification of fixed points in the space of all independent 
couplings represents a very interesting but rather involved problem, which 
will not be addressed here in full generality.  Instead, we will limit the 
number of independent couplings in \eetwelve\ by imposing further symmetry
on the theory.  

Note that in the component form of the general Lagrangian \eetwelve\ with  
arbitrary choice of the bare couplings, $B^\mu$ always comes out as a Gaussian 
auxiliary field, and can be integrated out exactly.  Indeed, neither higher 
powers of $B^\mu$ nor its world-sheet derivatives can emerge in the 
Lagrangian, since only the first derivatives of $X^\mu$ are allowed by 
dimension counting; in other words, the higher powers of $B^\mu$ as well as 
its derivatives are always irrelevant.  

\subsec{Theory with Local Parisi-Sourlas Supersymmetry}

So far we have discussed a theory whose symmetry is given by the semi-rigid 
Parisi-Sourlas supersymmetry \eesix , \eeseven .    
In this Section we go one step further, and enhance the gauge symmetries so 
that they will permit only one marginal term (and no relevant 
terms) in the world-sheet Lagrangian.  Upon partial gauge fixing, this theory
 with enlarged supersymmetry will belong to the class of theories \eetwelve\  
discussed in Section~4.1. 

We extend the supersymmetries \eesix , \eeseven\ to all super-diffeomorphisms 
of the super world-sheet $(\sigma^a,\theta^\alpha)$.  
In order to keep in contact with the general class of theories \eetwelve , 
we will use the following trick.  {}First, extend the space-time to a 
supermanifold by introducing two fermionic dimensions, with coordinates 
$\vartheta^m$, $m=1,2$ (we will also use the simpler notation 
$\vartheta^1\equiv\vartheta$ and $\vartheta^2\equiv\bar\vartheta$).  
The 
space-time manifold is now parametrized by
\eqn\eeaa{y^M\equiv(x^\mu,\vartheta^m).}
The flat space-time metric $G_{\mu\nu}=\delta_{\mu\nu}$ is extended to a 
super-metric $G_{MN}=(\delta_{\mu\nu},\varepsilon_{mn})$.  The embedding of 
the world-sheet to the space-time is described by world-sheet superfields, 
which we will denote by capital letters:
\eqn\eeaaa{Y^M\equiv(X^\mu,\Theta^m).}
The component expansion of $X^\mu$ is that of \eethree .

We require that the theory be gauge invariant under all world-sheet 
superdiffeomorphisms, and globally invariant under the space-time 
super-Poincar\'e symmetry $I\OSp(D|2)$.  The simplest Lagrangian with these 
symmetries is 
\eqn\eeab{\CL=\frac{1}{\alpha_0}\int_\Sigma d^2\sigma d^2\theta\sqrt{\sdet 
H_{AB}},}
where $A,B$ goes through all four indices of the supersymmetric world-sheet, 
and $\sdet H_{AB}$ is the superdeterminant of the induced supermetric 
\eqn\eeac{H_{AB}\equiv (-1)^{B(B+N)}\p_AY^M\,G_{MN}\,\p_BY^N.}
No relevant terms and no other marginal terms are allowed by local world-sheet 
supersymmetry and space-time super-Poincar\'e invariance.  

This theory indeed fits into the general class \eetwelve , once we partially 
fix the gauge so that the remaining gauge symmetry is the semirigid 
supersymmetry of \eesix , \eeseven .  To get a theory invariant under 
semirigid diffeomorphisms, we can fix the fermionic part of the gauge symmetry 
by imposing a Monge gauge in the fermionic dimensions:
\eqn\eead{\Theta=\theta,\qquad\qquad \bar\Theta=\bar\theta.}
This gauge fixing reduces the superdeterminant of the induced supermetric 
to 
\eqn\eeae{\sdet H_{AB}=\det H_{ab}\left(\frac{1}{1-S}\right)^2,}
and the Lagrangian \eeac\ becomes a particular example of a theory from the 
general class \eetwelve , with all coupling constants $g_{i0}$ fixed uniquely 
in terms of one coupling, $\alpha_0$.  

It is interesting to notice that in the theory with fully local Parisi-Sourlas 
supersymmetry, the free-field fixed point at small $\alpha_0$ does not fix the 
canonical dimension of the superfields $X^\mu,\Theta^m$.  The full Lagrangian 
\eeab\ before gauge fixing is invariant under the space-time scale 
transformations $Y^M\rightarrow\lambda Y^M$ that act trivially on the 
world-sheet.  This additional symmetry is not respected by the gauge fixing, 
which implicitly assumes $[\Theta]=[\theta]=-1$.  We could keep track of this 
symmetry during the gauge fixing by introducing a dimensionful parameter 
$\gamma$, and rewriting the Monge gauge as $\Theta=\gamma\theta$, $\bar\Theta
=\gamma\,\bar\theta$.  This parameter $\gamma$ has the dimension of space-time 
length and world-sheet mass.  

Another interesting fact about the fermionic Monge gauge is that it identifies 
translations along the fermionic dimensions on the world-sheet with 
supertranslations in the extended space-time.  In the topological 
interpretation of the world-sheet Lagrangian, world-sheet supertranslations 
correspond to the topological BRST (and anti-BRST) symmetry \eesusyalg .   

\subsec{Asymptotic Freedom}

The symmetries of our theory permit only one marginal coupling, the string 
rigitity $\alpha_0$.  In this section we will consider the weak-coupling 
expansion in $\alpha_0$, and calculate the one-loop beta function.  We will 
closely follow the spirit of \nelsonpowers\ and in particular \clnp , where a 
detailed analysis of the bosonic rigid string path integral has been 
performed.  Some additional information on the bosonic case can also be 
found in \refs{\jerusalem,\david,\pisarski}.  

It will be convenient to work in the off-shell, superfield formalism developed 
in Section~4.1.  Split the fields into a reference background plus 
fluctuations, $Y^M=\bar Y^M+\sqrt{\alpha_0}\delta  Y^M$.  We first keep 
$\bar Y^M$ arbitrary, but eventually we will evaluate the one-loop path 
integral for a conveniently chosen $\bar Y^M$.  

Choose normal gauge with respect to the reference surface $\bar Y^M$, 
\eqn\eepeffgauge{\p_A\bar Y\cdot\delta Y=0.}

The first contribution to the effective action will come from the determinant 
of the operator in the quadratic expansion of the Lagrangian in 
$\delta Y^M$'s, 
\eqn\eepeffone{\frac{1}{\alpha_0}\int_\Sigma d^2\sigma d^2\theta 
\sqrt{\sdet H_{AB}(\bar Y)}+\int_\Sigma d^2\sigma d^2\theta \,
\delta Y^M\CO_{MN}\delta Y^N.}
Here $\CO$ is the second-order operator density on the world-sheet, 
essentially the super-Laplace operator on the normal bundle in the reference 
background $\bar Y^M$. 

There are several other contributions to the effective action, and we have to 
investigate whether they can contribute to the one-loop beta function.  

First, the Faddeev-Popov determinant corresponding to our choice of gauge 
is not trivial \clnp .  Still, this determinant is independent of $\alpha_0$, 
and to first order in $\alpha_0$, we can simply evaluate it in the reference 
background $\bar Y^M$.  In this background, the determinant is given by the  
superdeterminant of the world-sheet supermetric,  
\eqn\eepefftwo{\Delta _{\rm FP}\sim (\sdet H_{AB}(\bar Y))^{-1}.}
Since the result is ultralocal in $(\sigma^a,\theta^\alpha)$, it will not 
contribute to the renormalization of $\alpha_0$.  All fluctuations will be of 
higher order in $\alpha_0$, and thus the FP determinant will not contribute to 
the one-loop beta function.  

The last contribution to the effective action comes from the careful 
analysis of the measure.  In general, the measure is non-linear, and $\delta 
Y^M$-dependent.  One can replace the full non-linear measure $[d\,\delta Y]$ 
by a flat measure $[d\,\delta Y]_0$,  defined with respect to some flat 
reference metric on the world-sheet.  This changes the density of degrees of 
freedom, which has to be compensated by a renormalization factor in the 
measure, 
\eqn\eepeffthree{[d\,\delta Y]e^{-\int\CL}=[d\,\delta Y]_0
e^{-\int\CL-\int\delta\CL}.}
Being a renormalization factor, this Liouville term $\delta\CL$ depends on 
the parameters of $\CL$.  

{}For the purpose of our gaussian integration, however, we do not need to know 
the exact form of the Liouville term.  In particular, we do not need the 
measure, just the metric on the space of fields.  The metric given by the 
full non-linear measure $[d \delta Y]$ coincides with the metric defined with 
respect to the fixed reference background $\bar Y^M$.  Thus, we do not have 
to worry about the Liouville term in our one-loop calculation of the 
beta function.  

The terms in the effective action that contribute to the renormalization of 
$\alpha_0$ are thus
\eqn\eeacteff{\CL_{eff}=\frac{1}{\alpha_0}\int_\Sigma d^2\sigma d^2\theta 
\sqrt{\sdet H_{AB}(\bar Y)}+\frac{1}{2}\,\str_{\bar H}\ln \CO ,}
where $\str_{\bar H}$ is the supertrace with respect to the background 
supermetric, $\str_{\bar H}\ldots=\int_\Sigma d^2\sigma d^2\theta \sqrt{\sdet 
H_{AB}(\bar Y)}\ldots$.

Having identified all possible contributions to the one-loop beta function of 
$\alpha_0$, we identify the renormalized coupling $\alpha$ as the inverse of 
the coefficient in front of $\int_\Sigma d^2\sigma d^2\theta
\sqrt{\sdet H_{AB}(\bar Y)}$ in the effective action.  As we have argued, 
the $\str_{\bar H}\ln\CO$ term is the only term that contributes to the 
renormalization of $\alpha_0$. 

So far our argument has been independent of the particular choice of 
$\bar Y^M$.  Now we will calculate the effective coupling 
by evaluating the contributions in a convenient background.  Notice that the 
area term $\int_\Sigma d^2\sigma d^2\theta\sqrt{\sdet H_{AB}(\bar Y)}$ has 
already been factorized from our definition of the renormalized coupling 
$\alpha$.  This allows us to conveniently calculate $\alpha$ by 
choosing for $\bar Y^M$ the flat super world-sheet, thus obtaining
\eqn\eepeffdet{\frac{1}{\alpha(\Lambda)}=\frac{1}{\alpha(\Lambda_0)}+
\frac{D-2}{2}\int_{\Lambda_0}^\Lambda\frac{d^2k\,d^2\kappa}{(2\pi)^2}\ln 
(k^2-2\kappa\bar\kappa), }
where $k_a,\kappa_\alpha$ are the Fourier super-momenta on the flat super 
world-sheet.%
\foot{{}For the super Fourier transform on Parisi-Sourlas superspace, see 
e.g.\ \mitstoch .} 
We can evaluate the momentum integral using the superfield expansion of the 
logarithm, 
\eqn\eeeval{\int_{\Lambda_0}^\Lambda\frac{d^2k\,d^2\kappa}{(2\pi)^2}\ln (k^2-
2\kappa\bar\kappa)=\int_{\Lambda_0}^\Lambda\frac{d^2k\,d^2\kappa}{(2\pi)^2}
\left(\ln k^2-2\frac{\kappa\bar\kappa}{k^2}\right)=-\frac{1}{\pi}\ln (\Lambda/
\Lambda_0).}
Notice that our superspace calculation is a calculation with a supersymmetric 
cutoff.  In the integration over the shell in the momentum space, the momentum 
modes of the four components of the superfield $Y^M$ are all cut off at 
the same lower and upper scales $\Lambda_0$ and $\Lambda$.  

To first order in $\alpha_0$, \eepeffdet\ and \eeeval\ gives the effective 
coupling $\alpha (p)$ 
\eqn\eepeffbeta{\alpha(p)=\frac{\alpha_0}{1-\frac{D-2}{4\pi}
\alpha_0\,\ln\left(\frac{\Lambda^2}{p^2}\right)}.}
Thus, we have demonstrated to order $\alpha_0$ that the theory is 
asymptotically free in dimensions $D>2$, and $D=2$ is the critical dimension.  

\subsec{Quantum Theory at Large $D$}

We have seen that in dimensions greater than two, the theory is 
asymptotically free.  Non-perturbative effects in $\alpha_0$ will be 
very important in the long-distance dynamics of the theory.  To shed some 
light on the non-perturbative behavior of the model, we will analyze the 
theory in the mean field approximation at large number of space-time 
dimensions.  Large-$D$ behavior of the bosonic rigid string was studied in 
\refs{\pisarski,\david}.  

The Lagrangian that will be convenient to use in the $1/D$ expansion is 
\eqn\eeda{\eqalign{\CL&=\frac{1}{\alpha_0}\int_\Sigma d^2\sigma d^2\theta
\sqrt{\sdet H_{AB}}\cr
&\qquad{}-\frac{i}{\alpha_0}\int_\Sigma d^2\sigma d^2\theta\lambda^{AB}
\left\{H_{AB}-(-1)^{B(B+N)}\p_AY^MG_{MN}\p_BY^N\right\}.\cr}}
Now we treat $H_{AB}$ as an independent supermetric on the world-sheet, and 
$\lambda^{AB}$ is the Lagrange multiplier that identifies $H_{AB}$ classically 
with the induced supermetric.  

The effective action of the model in the mean field approximation at large 
$D$ can be calculated as follows.  We again split $Y^M$ into the classical 
background part and a quantum part, 
\eqn\eedb{Y^M=Y^M_0+\sqrt{\alpha_0}Y^M_{qu}.}
The Lagrangian is quadratic in $Y^M$ and we can integrate over $Y^M_{qu}$ 
exactly.  This gives 
\eqn\eedc{\eqalign{\CL_{\rm eff}&=\frac{1}{\alpha_0}\int_\Sigma d^2\sigma 
d^2\theta\sqrt{\det H_{AB}}\cr
&\qquad{}-\frac{i}{\alpha_0}\int_\Sigma d^2\sigma d^2\theta\lambda^{AB}
\left\{H_{AB}-(-1)^{B(B+N)}\p_AY^M_0G_{MN}\p_BY^N_0\right\}\cr
&\qquad{}+\frac{1}{2}\str_H\ln\,(\frac{i}{\sqrt{\sdet H_{AB}}}\p_A
\lambda^{BA}\p_B).\cr}}
We still have $D$ background fields $Y_0^M$ explicitly in the Lagrangian, 
but since they only enter through their induced metric 
$H_{AB}(Y_0)=\p_AY^M_0G_{MN}\p_BY^N_0$, they do not spoil the large-$D$ 
limit.  

The equations of motion following from \eedc\ are 
\eqn\eedd{\eqalign{\sqrt{\sdet H_{CD}}\;H^{BA}&=-i\lambda^{AB},\cr
H_{AB}-H_{AB}(Y_0)=\alpha_0\frac{\delta}{\delta\lambda^{AB}}\str_H
\ln&\left(\frac{i}{\sqrt{\sdet H_{EF}}}\p_C\lambda^{DC}\p_D\right),\cr
\p_A(\lambda^{BA}\p_BY_0^M)&=0.\cr}}
To find a solution, consider first an infinite planar world-sheet in 
space-time.  We can expect the classsical solution $\lambda^{AB}$ and 
$H_{AB}$ to be independent of $\sigma^a$.  It might be tempting to assume 
their independence of $\theta^\alpha$ as well.  This would lead to the 
following Ansatz:
\eqn\eedg{H_{AB}(\sigma^a,\theta^\alpha)=\omega\;H_{AB}(Y_0)= \omega\;
H^{(0)}_{AB},}
with $\omega$ a ($\theta^\alpha$-independent) constant.  Here $H^{(0)}_{AB}$ 
is the flat supermetric 
\eqn\eedh{H^{(0)}_{ab}=\delta_{ab},\quad H^{(0)}_{\alpha\beta}\equiv
\epsilon_{\alpha\beta},\quad H^{(0)}_{a\alpha}= H^{(0)}_{\alpha a}= 0.}
Plugging this Ansatz into the mean-field equations, we can see immediately 
that the assumption of $\theta^\alpha$-independence cannot be correct; this 
Ansatz gives no saddle point.  This may have been guessed even before any 
calculation, by simply noting that the Ansatz does not contain any 
dimensionful parameters, which makes it insufficient for dimensional 
transmutation that we expect in the large $D$ expansion on the basis of 
asymptotic freedom.  

The Ansatz must be modified, and precise details will depend on the boundary 
conditions imposed on the world-sheet.  It is natural to consider
\eqn\eenewans{H_{ab}=(\omega+(\mu+\tau)\,\theta\bar\theta)\,\delta_{ab},\qquad
H_{\alpha\beta}=(\omega+(\mu-\tau)\,\theta\bar\theta)\,\epsilon_{\alpha\beta},}
with $\tau$ and $\mu$ parameters to be determined by the saddle-point 
condition \eedd .  The dimensionful parameter $\tau$ plays the role of 
effective string tension,  
\eqn\eeitstension{\int_\Sigma d^2\sigma d^2\theta \sqrt{\sdet H_{AB}}=
\tau\int_\Sigma d^2\sigma d^2\theta \sqrt{h}\theta\bar\theta=\tau\int_\Sigma 
d^2\sigma \sqrt{h}.}
One can easily see that for non-zero $\tau$, the modified Ansatz \eenewans\ 
does lead to a saddle point, with the saddle-point value of $\tau$ determined 
via 
\eqn\eemudet{\tau\sim \Lambda^2 \exp\left(-\frac{4\pi}{D\alpha_0}\right).}
This is a standard result of dimensional transmutation in an asymptotically 
free theory.  

Note that the dynamically generated tension $\tau$ breaks the Parisi-Sourlas 
supersymmetry on the world-sheet.  Indeed, the presence of the 
$\theta$-dependent term in \eenewans\ means that the saddle-point value of 
$H_{AB}$ is not invariant under the scalar Parisi-Sourlas supersymmetries 
$Q$ and $\bar Q$ of \eepsalg .  Unless the theory flows into a non-trivial 
infrared fixed point, we should expect that the long-distance behavior will 
be dominated by the tension term, dynamically generated by dimensional 
transmutation.  The Parisi-Sourlas supersymmetry is only restored 
asymptotically in the ultraviolet, where it effectively plays 
the role of topological symmetry.  

So far we have studied the large-$D$ theory on an infinite plane $\R^2$, 
and have argued that dimensional transmutation generates a non-zero string 
tension.   In order to obtain any information about the string spectrum, one 
should study the theory on a world-sheet of topology $\R^1\times\S^1$, 
which correspond to the free closed string.  Asymptotic freedom and 
dimensional transmutation imply that the ground state of the closed string 
is described by the full non-perturbative vacuum of the world-sheet 
field theory, on a world-sheet with non-trivial topology $\R^1\times\S^1$.  

Alternatively, one can consider the following space-time point of view.  
In elementary closed string theory, where we have linear Regge trajectories, 
the ground state of the closed string is just the perturbative Fock vacuum, 
with at most a finite number of world-sheet oscillators excited.  In the 
classical bosonic rigid string, the non-zero string rigidity prevents the 
classical ground state%
\foot{For the purpose of this argument, we can ignore the perturbative 
instability of the classical bosonic rigid string.}
from collapsing into a point, the ground state becomes a hoop of non-zero 
radius, and the Regge trajectories are consequently bent at low masses 
\pisarski .  In our supersymmetric model, however, this does not happen 
classically on the world-sheet.  Since there is no length scale in the 
classical world-sheet 
Lagrangian, there is nothing that would fix the radius of the hoop.  
Classically, there is no tension, and the string rigidity decompactifies the 
classical world-sheet.  In the quantum theory, on the other hand, the length 
scale generated by dimensional transmutation gives a non-zero tension to the 
string, and the radius (of extrinsic curvature) of the string in its ground 
state is set dynamically. 

One very important issue which is not entirely clear concerns unitarity of the 
theory.  On the face of it, we seem to have two sources of possible violations 
of unitarity.  The Lagrangian is of higher order in derivatives -- a property 
shared by the bosonic rigid string, where it leads to ghost states at least 
in perturbation theory at small $\alpha_0$.  In addition to that, our 
Parisi-Sourlas supersymmetric theory has fields with the ``wrong'' statistics, 
namely the fermions $\psi^\mu$ and $\chi^\mu$, which represent another 
possible source of ghost states.  However, things may not really be as bad as 
they seem.  Due to the local Parisi-Sourlas supersymmetry, gauge invariant 
states satisfy $Q\left|{\rm phys}\right\rangle =0$.  This means that as as 
long as supersymmetry is unbroken, which is the case asymptotically in the 
ultraviolet, the theory is effectively a topological string theory, and 
unitarity should not be violated.  We have argued that due to the asymptotic 
freedom of the theory, world-sheet supersymmetry is broken at all scales, and 
is only asymptotically restored in the UV limit.  At long distances, the 
dynamically generated tension term dominates (assuming that the theory does 
not flow into a non-trivial IR fixed point), and the string is described by an 
effective Nambu-Goto Lagrangian.  In this sense, the full supersymmetric rigid 
string theory provides an ultraviolet regularization of the Nambu-Goto string 
theory, in a way compatible with the apparent absence of conventional stringy 
degrees of freedom in the ultraviolet.   

\subsec{String Tension in Two Dimensions and the Superpotential}

In two space-time dimensions, world-sheet supersymmetry stays unbroken at all 
scales, non-zero tension is not dynamically generated, and we need another 
mechanism for incorporating non-zero tension in the theory.  This can be 
done if we relax the global symmetry restrictions and permit soft breaking 
of the full space-time super-Poincar\'e symmetry $I\OSp(2|2)$ to its 
bosonic, Poincar\'e subgroup $ISO(2)$.  

Once we relax the condition of space-time $I\OSp(2|2)$ symmetry, we can write 
down a ``superpotential'' term 
\eqn\eeba{\delta\CL=\tau_0\int_\Sigma d^2\sigma d^2\theta\sqrt{\sdet H_{AB}}
\Theta\bar\Theta.}
This term of course respects the local Parisi-Sourlas supersymmetry on the 
world-sheet.  The coupling constant $\tau_0$ has canonical dimension two in 
mass units, and will play the role of the string tension as we are just about 
to see.    

In order to make contact with the topological string formulation of 
Section~3.1b, we want to partially fix the gauge by choosing the fermionic 
Monge gauge \eead\ again.  In this gauge, the remaining symmetries are those 
of the topological rigid string of Section~3.1, and the superpotential becomes
\eqn\eebb{\delta\CL=\tau_0\int_\Sigma d^2\sigma\sqrt h(1+S_0+\ldots).}
Here $S_0$ is the $\theta=\bar\theta=0$ component of $S$.  In terms of 
the component fields, $S_0$ is equal to the scalar product of the normal 
components of $\psi$ and $\chi$:  
\eqn\eebc{S_0=\psi\cdot\chi-\psi\cdot\p_a x\;h^{ab}\;\p_b x\cdot\chi.}
Thus, $\tau_0$ behaves like the string tension, and the superpotential is 
equivalent to the Nambu-Goto Lagrangian, improved by a finite number of 
fermionic corrections.  

This term precisely corresponds to the tension term in \srst , eq.\ (6.21). 
It is interesting 
to notice that in the fully supersymmetric theory before gauge fixing, this 
term respects both local world-sheet supersymmetry and space-time Poincar\'e 
symmetry.  After the partial gauge fixing, however, this term explicitly 
breaks the residual global fermionic symmetry that corresponds to the 
topological BRST charge in the formal topological interpretation of the 
theory.  This explains why the non-zero tension case was somewhat 
difficult to understand in the conventional framework of two-dimensional 
topological rigid string theory \srst .  Even in two space-time dimensions, 
it is natural to consider the more general framework of rigid string 
theory with local Parisi-Sourlas supersymmetry.  

\bigskip\medskip\noindent
I benefited from useful discussions with Tom Banks, Mike Douglas, David Gross, 
Clifford Johnson, and Sasha Polyakov.  I wish to thank the Aspen Center for 
Physics and the Rutgers High-Energy Theory Group for their generous 
hospitality and stimulating atmosphere during the initial and final stages of 
this work, respectively.  This work has been supported by a Sherman Fairchild 
Prize Fellowship and by DOE grant DE-FG03-92-ER~40701. 

\listrefs
\end